\documentclass[sn-mathphys,Numbered]{sn-jnl}
\usepackage{graphicx}%
\usepackage{multirow}%
\usepackage{amsmath,amssymb,amsfonts}%
\usepackage{amsthm}%
\usepackage{mathrsfs}%
\usepackage[title]{appendix}%
\usepackage{xcolor}%
\usepackage{textcomp}%
\usepackage{manyfoot}%
\usepackage{booktabs}%
\usepackage{algorithm}%
\usepackage{algorithmicx}%
\usepackage{algpseudocode}%
\usepackage{listings}%
\usepackage{enumitem}
\theoremstyle{thmstyleone}%
\newtheorem{theorem}{Theorem}
\newtheorem{proposition}[theorem]{Proposition}%
\newtheorem{lemma}[theorem]{Lemma}
\newtheorem{corollary}{Corollary}[theorem]
\theoremstyle{thmstyletwo}%
\newtheorem{example}{Example}%
\newtheorem{remark}{Remark}%
\theoremstyle{thmstylethree}%
\newtheorem{definition}{Definition}%

\newtheorem{problem}{Problem}%
\newtheorem{observation}{Observation}
\RequirePackage{amsthm}
\usepackage{graphicx}%
\usepackage{multirow}%
\usepackage{amssymb,amsfonts}%
\usepackage{amsmath}
\usepackage{mathrsfs}%
\usepackage[title]{appendix}%
\usepackage{xcolor}%
\usepackage{textcomp}%
\usepackage{manyfoot}%
\usepackage{booktabs}%
\usepackage{algorithm}%
\usepackage{algorithmicx}%
\usepackage{algpseudocode}%
\usepackage{listings}%
\usepackage{comment}
\usepackage{caption}
\usepackage{float}
\usepackage{multirow}
\usepackage{tabularx}
\usepackage{hyperref}
\usepackage[T1]{fontenc}

\usepackage{booktabs}
\usepackage{enumitem} 

\newcommand{\di}{\mathrm{diam}}

\begin{document}

\title[Binary Stretch Embedding of Weighted Graphs]{Binary Stretch Embedding of Weighted Graphs}

\author*[1]{\fnm{Javad} \sur{B. Ebrahimi}}\email{	javad.ebrahimi@sharif.edu}

\author[1]{\fnm{Mehri} \sur{Oghbaei Bonab}}\email{m.oghbaei95@sharif.edu}

\affil[1]{\orgdiv{Department of Mathematical Sciences}, \orgname{Sharif University of Technology}, \orgaddress{\street{Azadi Avenue}, \city{Tehran}, \postcode{11155-9415}, \state{Tehran}, \country{Iran}}}


\abstract{The abstract}
\abstract{

 In this paper, we introduce and study the problem of \textit{binary stretch embedding} of edge-weighted graph. This problem is closely related to the well-known \textit{addressing problem} of Graham and Pollak. Addressing problem is the problem of assigning the shortest possible length strings (called ``addresses") over the alphabet $\{0,1,*\}$ to the vertices of an input graph $G$ with the following property. For every pair $u,v$ of vertices, the number of positions in which one of their addresses is $1$, and the other is $0$ is exactly equal to the distance of $u,v$ in graph $G$. When the addresses do not contain the symbol $*$, the problem is called \textit{isometric hypercube embedding}. As far as we know, the isometric hypercube embedding was introduced by Firsov in 1965. It is known that such addresses do not exist for general graphs.
 
 Inspired by the addressing problem, in this paper, we introduce the \textit{binary stretch embedding problem}, or BSEP for short, for the edge-weighted undirected graphs. We also argue how this problem is related to other graph embedding problems in the literature. 

 Using tools and techniques such as Hadamard codes and the theory of linear programming, several upper and lower bounds as well as exact solutions for certain classes of graphs will be discovered.
 
 As an application of the results in this paper, we derive improved upper bounds or exact values for the maximum size of  Lee metric codes of certain parameters.
 }




\maketitle
\section{Introduction}\label{sec1}
 In this paper, we introduce and study the problem of \textit{binary stretch embedding} of an edge-weighted graph. This problem is closely related to the well-known \textit{addressing problem} of Graham and Pollak. We utilize the result of this study to derive novel bounds for Lee metric codes of certain parameters. 
 
 In the continuation, we describe both of these problems as well as Lee metric codes in detail. 
 
Consider the set of $n$-tuples from the alphabet set $\{0, 1, *\}$ (called ``addresses''). The distance between two addresses is defined as the number of places where one has a $0$ and the other a $1$ (Thus, the stars do not contribute to the distance). An addressing of a graph $G$ is an assignment of addresses to the vertices such that the distance of any two vertices in $G$ is equal to the distance of their addresses. For a given graph $G$, what is the smallest possible $n$ such that an addressing of length $n$ exists? This number is called the addressing number of $G$ and is denoted by $N(G)$. In \cite{grah}, the addressing problem is presented and defined as part of a switching theory application.

Moreover, the addressing number of graph $G$ is equal to the biclique partition number of its distance multigraph. This is shown by Graham and Pollak in \cite{grah}, \cite{grah2}. The biclique partition number of a graph $G$, denoted by $bp(G)$, is the minimum number of complete bipartite subgraphs (bicliques) of $G$ that cover all the edges of $G$.
The distance multigraph of $G$, denoted by $D(G)$, is a multigraph with the same vertices of $G$. In $D(G)$, the multiplicity of any edge $e=uv$ is the distance between $u$ and $v$ in $G$.
\\
One can observe that the defined notion of distance is not a metric on the set $\{0, 1, *\}$. 
Thus, it is natural to ask the same problem for the binary alphabet equipped with the Hamming distance. In fact, for the binary alphabet, the problem becomes the following question. What is the smallest integer $l$ such that there exists an isometry (i.e. distance preserving map) $\phi: \big(V(G),d_G\big) \to \big(\{0,1\}^l, d_H\big)$? Here, $d_G$ is the distance function on $G$ with respect to the weight of edges. Also, $d_H$ is the usual Hamming distance (See \ref{codingth} for the definition). This problem is closely related to the ``isometric hypercube embedding'' problem, which only asks about the existence of any $\phi$, regardless of the length $l$. 

As far as we know, this problem was first introduced in the paper \cite{iso} by Firsov. Isometric embedding into hypercubes is further studied in \cite{distanceperserv}, \cite{iso2}, \cite{scale}. Moreover, in \cite{wilk} and \cite{wink}, isometric Hamming embedding of unweighted graphs is investigated. The book \cite{dezabook} by Deza and Laurent, and the notes \cite{matouseknotes}  and \cite{chepoinotes}, by Matousek and Chepoi are great references to isometric embedding and metric embedding in general. 

Isometric hypercube embedding problem has applications in the design of DNA strands \cite{dna}, communication, and coding theory \cite{iso}, \cite{appli2}. For weighted graphs, isometric Hamming embedding has been studied recently in \cite{weighted}.
\\
It has been shown in \cite{scale}, \cite{football} that there exist graphs that do not admit any isometric embedding into hypercubes, no matter how large we pick $l$. For instance, one can easily show that there is no addressing for $K_3$ over the binary alphabet, where $K_3$ is a complete graph with $3$ vertices. However, it is still a legitimate question to ask for the smallest length $l$ such that there exists a function $\phi: \big(V(G),d\big) \to \big(\{0,1\}^l, d_H\big)$ such that for every pair $u,v$ of the vertices, $d(u,v) \leq d_H\big(\phi(u),\phi(v)\big)$. This problem is introduced and studied in the current paper. In Section \ref{formulation}, the precise formulation of this problem as well as a related version of it has been described 

These types of questions are studied in the field of metric embedding of graphs, which has also received significant attention, primarily due to their applications in computer science and algorithm design. (See \cite{embed1}, \cite{embed2}.)\\

We can take one further step by considering edge-weighted graphs. For a connected weighted graph $G$ with a weight function $\mathbf{w}$, we can still ask for the minimum length of binary addressing. This means, for all $u,v \in V(G)$ we have $\phi : V(G) \to \{0,1\}^l$, such that $d_\mathbf{w}(u,v) \leq d_H\big(\phi(u),\phi(v)\big)$. Here, $d_\mathbf{w}(u,v)$ is the weighted distance of $u,v$. The smallest such $l$ is denoted by $c(G,\mathbf{w})$.
\\
Note that whenever the graph has $n$ vertices and the weight function is constant $d$, our problem reduces to the following well-known problem in the context of error-correcting codes: 
What is the minimum length length binary code, containing $n$ codewords, and, minimum distance $d$? 
\\
We also define $c_{\lambda}(G,\mathbf{w})$ as $c(G, \mathbf{\lambda}\mathbf{.w})$ in which $\mathbf{\lambda}\mathbf{.w}$ is a weight function which assigns the weight $\lambda. w(e)$ to the edge $e$. Here, $\lambda$ is a positive integer scalar. Observe that in this notation, $c_1(G,\mathbf{w})=c(G,\mathbf{w})$.
\\
Unlike isometric embedding, which does not necessarily exist for arbitrary graphs, function $\phi$ exists for every connected weighted graph $G$, provided that the length $l$ is large enough. Thus, $c_\lambda(G,\mathbf{w})$ is well defined for any connected graph $G$, positive integer weight $\mathbf{w}$ and positive integer scalar $\lambda$. The main challenge is to minimize the size of the corresponding vectors (which we call the \textit{binary addresses}).
\\
In Table \ref{comparetable}, we compare several graph embedding-type problems to those we introduce in this paper.

\subsection{Informal Statement of Our Results}
Before going into the technical details, we informally outline the main results of this paper. 
\begin{itemize}
\item{\textbf{General upper and lower bounds and some exact results:}} 
\\
 In Lemma \ref{threepart}, we derive lower and upper bounds on $c(G,\mathbf{w})$ for arbitrary edge-weighted graph $G$ in terms of other graph parameters. This, in turn, provides an exact formula for $c(G,\mathbf{w})$ for certain graphs. This result is formally stated and proved in Theorem \ref{cyc}.

    \item {\textbf{Integer programming formulation and its linear relaxation:}} 
    \\
   In \ref{lp}, we formulate the problem as an integer program. We consider the linear relaxation problem and present the relationship between the relaxed problem and $\lambda$-BSEP. Corollary \ref{mu} contains this result.
   
   In Theorem \ref{plot}, by using the weak duality theorem, we derive a lower bound for the relaxed problem. This, in turn, implies a stronger lower bound for $c(G,\mathbf{w})$. 
   
   By utilizing Hadamard codes, we find an upper bound at most twice the lower bound. This will provide a 2-approximation solution for the linear problem. Corollary \ref{beta} summarizes this result. This result helps us to find the integrality gap of the relaxed problem. 
  
     \item{\textbf{BSEP for the Cartesian product of graphs:}} 
    \\
     Using the results in the previous parts and based on the Cartesian product of graphs, we construct a family of graphs with the exact solutions to BSEP or its linear relaxation counterpart. 
    \item{\textbf{Application to Lee metric codes:}}
    \\
      We present an application of our results in the context of Lee metric codes. As a theorem, we obtain a relation between the size of Lee codewords and binary error correction codes. This will help us to employ our findings to improve the bounds on the size of Lee metric codes of certain parameters. Table \ref{leetable} summarises the improved bounds. 
\end{itemize}
\subsection{Organization of the Paper}

In the rest of this paper, we start with an overview of the preliminaries. This includes the basic definitions and notations of coding theory, graph theory, and linear programming which we use in the subsequent sections. 
\\
In Section \ref{formulation}, we formally define the BSEP and $\lambda$-BSEP for edge-weighted graphs. 
\\
Next, we present the main results of the paper. Finally, as an application of our result, we present new bounds for Lee metric codes of certain parameters.\\
In Appendix A, we present an alternative proof for one of the theorems as the Plotkin type bound. 

Appendix B contains a table that summarizes notations. In Appendix C, to gain a better understanding of the subject of this work, we compare some other graph embedding problems to those introduced in this paper. 
\section{Preliminaries}\label{sec2}
In this section, we review basic definitions and notations from coding theory, graph theory, and linear programming.
\subsection{Metric Space and Coding Theory Terminology}\label{codingth}
First, we define the metric space and then a particular case of mapping between metrics.
\begin{definition}[Metric space]
    A metric space is a set $M$ together with a distance function $d: M \times M \rightarrow \mathbb{R}_ {\geq 0}$ satisfying the following properties:
\begin{enumerate}
    \item 
   $ \forall \hspace{0.5em} x \in M, d(x,x)=0$
    \item
    $\text{If}\quad x\neq y, d(x,y)>0$
    \item
    $\forall \hspace{0.5em} x,y \in M, d(x,y)=d(y,x)$
    \item
     $\forall \hspace{0.5em} x,y,z \in M, d(x,z)\leq d(x,y)+d(y,z)$ (Triangle inequality).
\end{enumerate}
\end{definition}
There are different mappings between metric spaces. One of them is defined as follows. 
\begin{definition}[Stretch mapping]
    A mapping $\phi$ from metric space $(X,d)$ to another metric space $(Y, d^\prime)$ is stretch mapping when:
    $\forall \hspace{0.5em} a,b \in X, d(a,b)\le d^\prime\big(\phi(a),\phi(b)\big)$.
\end{definition}
Let $Q$ be a set of size $q$ and $d$ be some metric over $Q$. A code $C$ of length $n$ is a subset $C \subseteq Q^n$. When $q=2$, we call the code ``binary''. The elements of $C$ are called codewords. 
The metric $d$ naturally extends over the set $Q^n=\{(a_1, \dots, a_n): a_i \in Q, \hspace{0.5em} \forall 1\le i \le n\}$. We abuse the notation and denote the extended metric by $d$. One can easily check that the distance measure $d: Q^n \times Q^n \rightarrow \mathbb{R}_ {\geq 0}$ defined by $d\big((x_1, \dots, x_n), (y_1, \dots, y_n)\big):= \sum_{i=1}^{n} d(x_i, y_i)$ is a metric. The minimum distance of the code $C$ is defined as $\min_{\substack{{x,y \in Q^n}\\{x\neq y}}} d(x,y)$.
\\
An example of a metric space is when $M$ is a finite set and the distance function $d_H$ is defined as follows: $d_H(a,b)=1$ when $a\neq b$ else $0$. This metric space is called the Hamming space, and the function $d_H$ is called the Hamming distance. Another example of a metric space is the Lee metric, which is defined as follows. Take $Q=\{0, 1, \dots, q-1\}$ and $d(i,j)= \min \big\{|i - j|, q-|i - j|\big\}$. The intuition of the Lee metric is that if one arranges the numbers $0, 1, \dots, q-1$ cyclically around a cycle, the Lee distance of $i$ and $j$ is the length of the shorter arc these two vertices form. We define the concatenation of two codes as follows.
\begin{definition}[Concatenation]
    If $u=(u_1, \dots, u_m)$, $v=(v_1, \dots, v_n)$. Then the concatenation of $u$, $v$ denoted by $uv$ is defined as the length $m+n$ string $uv=(u_1, \dots, u_m, v_1, \dots, v_n)$.
\end{definition}\label{concat}
The Hadamard code is a set of $2^{m+1}$ codewords of length $2^m$ over the binary alphabet with the minimum distance of $2^{m-1}$.
\\
We denote the maximum size of a binary code of length $n$ and the minimum distance $d$ by $A_2(n,d)$. The Plotkin bound states that: If $2d>n$, then $A_2(n,d) \leq \frac{2d}{2d-n}$.
\\
Similarly, $A_q^L (n,d)$ denotes the maximum number of Lee codes over an alphabet set of size $q$, length $n$, and minimum distance of $d$. 
\subsection{Graph Theory Terminology}
We follow the notation of \cite{bondy} for graph theory basics. Let $G=(V, E)$ be a graph. A $(u,v)$-path is a path with endpoints $u,v$.
\\
An edge-weighted graph (or weighted graph) is a graph $G=(V, E)$ together with a weight function $\mathbf{w}: E \rightarrow \mathbb{R}$. In this paper, we only consider positive integer weights and those weighted graphs for which each edge is the shortest path between its
endpoints, which we call ``weight-minimal'' graphs (see \cite{weighted}). We denote a weighted graph $G$ with the weight function $\mathbf{w}$ by $(G,\mathbf{w})$ or $G_\mathbf{w}$. The weight of a subgraph $H$ of $(G,\mathbf{w})$ is the sum of the weights of the edges in $H$. For every connected weighted graph $(G,\mathbf{w})$, the smallest weight $(u,v)$-path is denoted by $d_\mathbf{w} (u,v)$ or $d_G (u,v)$ when there is no ambiguity about $\mathbf{w}$. The diameter of a weighted connected graph $(G,\mathbf{w})$ is defined as $\max_{u,v \in V} d_\mathbf{w} (u,v)$ and is denoted by $\di_\mathbf{w} (G)$. Moreover, we define $h_p(G_\mathbf{w}, \mathbf{w})$ and $h_c(G_\mathbf{w}, \mathbf{w})$ as the lowest weight of a Hamilton path and the lowest weight of a Hamilton cycle in the graph $G_\mathbf{w}$, respectively.   
\\
The Cartesian product $G \square H$ of two graphs $G$ and $H$, is a graph that has the vertex set as the Cartesian product $V(G) \times V(H)$. Also, two vertices $(u,u^\prime)$ and $(v,v^\prime)$ are connected in $G \square H$ if and only if either
$u = v$ and $u^\prime$ and $v^\prime$ are neighbors in $H$, or $u^\prime = v^\prime$ and $u$ and $v$ are neighbors in $G$.Also, graphs $G$, and $H$ are called factors. When two graphs are weighted such as $(G_1, \mathbf{w}_1)$ and $(G_2, \mathbf{w}_2)$, the Cartesian product of them is denoted by $(G_1 \square G_2, \mathbf{w}_1 \square \mathbf{w}_2)$. In Section \ref{cartpro}, by $G_1 \square G_2$ we mean $(G_1 \square G_2, \mathbf{w}_1 \square \mathbf{w}_2)$.
\subsection{Linear Programming Terminology}
In this paper, we consider the standard form of linear programs, namely:
\begin{equation*}
    \begin{array}{ll}
         \min & C^T X  \\
         \mathrm{s.t.} & 
         AX\geq b, \hspace{0.5em} X\geq 0.
    \end{array}
\end{equation*}
where $C$ is a vector defining the linear objective function, and $A$ is the constraint matrix. Also, $b$ is a vector. The feasible region of this problem is the set of all $X$ with $AX\geq b, X\geq 0$ (means all the entries of vector $X$ are non-negative). When the extra integrality condition on $X$ is imposed, the problem is called the integer program of the original linear program. The dual program will be the following problem:
\begin{equation*}
    \begin{array}{ll}
         \max & b^T Y  \\
         \mathrm{s.t.} & 
         A^TY\le C, \hspace{0.5em} Y\geq 0.
    \end{array}
\end{equation*}
One of the main results in this field is the weak duality theorem, which states: Let $x^*$
be a feasible solution to the primal and let $y^*$
be a feasible solution to the dual. Then, we have $b^T y^* \le C^T x^*$.
\section{Problem Formulation}\label{formulation}
In this section, we introduce the notion of binary address and present the problem formulation of BSEP and $\lambda$-BSEP.
\begin{definition}[BSEP]\label{parameterc}
A binary addressing of a weighted graph $(G,\mathbf{w})$ is a function $f: V(G) \rightarrow \{0,1\}^m$ in which $m$ is some positive integer number and for all $u,v \in V(G)$, we have $d_G (u,v) \le d_H \big(f(u), f(v)\big)$. (Here, $d_G$ and $d_H$ are the graph distance and the Hamming distance, respectively.) The number $m$ is called the length of the binary addressing of $f$, and $f(u)$ is referred to the binary address (or ``address'' for short) of the vertex $u$. BSEP is the problem of finding the minimum integer $m$ for which a binary addressing of length $m$ exists. This parameter is called the binary addressing number of $(G,\mathbf{w})$ and is denoted by $c(G,\mathbf{w})$.
\end{definition}
In other words, BSEP asks about the smallest hypercube graph such that we can embed the vertices of an edge-weighted graph to the vertices of the hypercube in such a way that the distances stretch. Alternatively, if we think of an edge-weighted graph as a discrete metric space, we would like to know the smallest size Hamming space such that there exists a metric stretching mapping from the graph to that space.
\begin{definition}[$\lambda$-BSEP]
For an edge-weighted graph $(G,\mathbf{w})$, and for the positive integer scalar $\lambda$, the $\lambda$-BSEP is the following question:

Find the value of $c_{\lambda}(G,\mathbf{w})$. Recall that $c_{\lambda}(G,\mathbf{w}) = c(G,\lambda .\mathbf{w})$ in which $\lambda .\mathbf{w}$ stands for the weight function obtained from $\lambda$ scaling of  $\mathbf{w}$ (hence the name $\lambda$-scaling) and $c(G,\mathbf{w})$ is defined in Definition \ref{parameterc}. 

\end{definition}
\begin{remark}
If we let $\mathbf{1}$ be the constant weight $\mathbf{1}$ on all the edges of a connected graph $G$, then we have $c(G, \mathbf{1}) \leq N(G)$.
\end{remark}
\begin{remark}
    Note that if one can answer BSEP for a graph $G$ and arbitrary weight $\mathbf{w}$, they can answer $\lambda$-BSEP as well. Thus, it might look bizarre to define $\lambda$-BSEP as a separate problem. The actual reason behind the definition of $\lambda$-BSEP, is that for a given weighted graph $(G,\mathbf{w})$, the asymptotic behavior of the answer to $\lambda$-BSEP, as $\lambda$ approaches infinity, is related to the fractional version of BSEP. We thoroughly cover this subject in section \ref{lambdasec}.
\end{remark}
\begin{example}
Consider the following weighted graph. One can assign these addresses of length $3$ to the vertices:
\begin{center}
\includegraphics[scale=0.25]{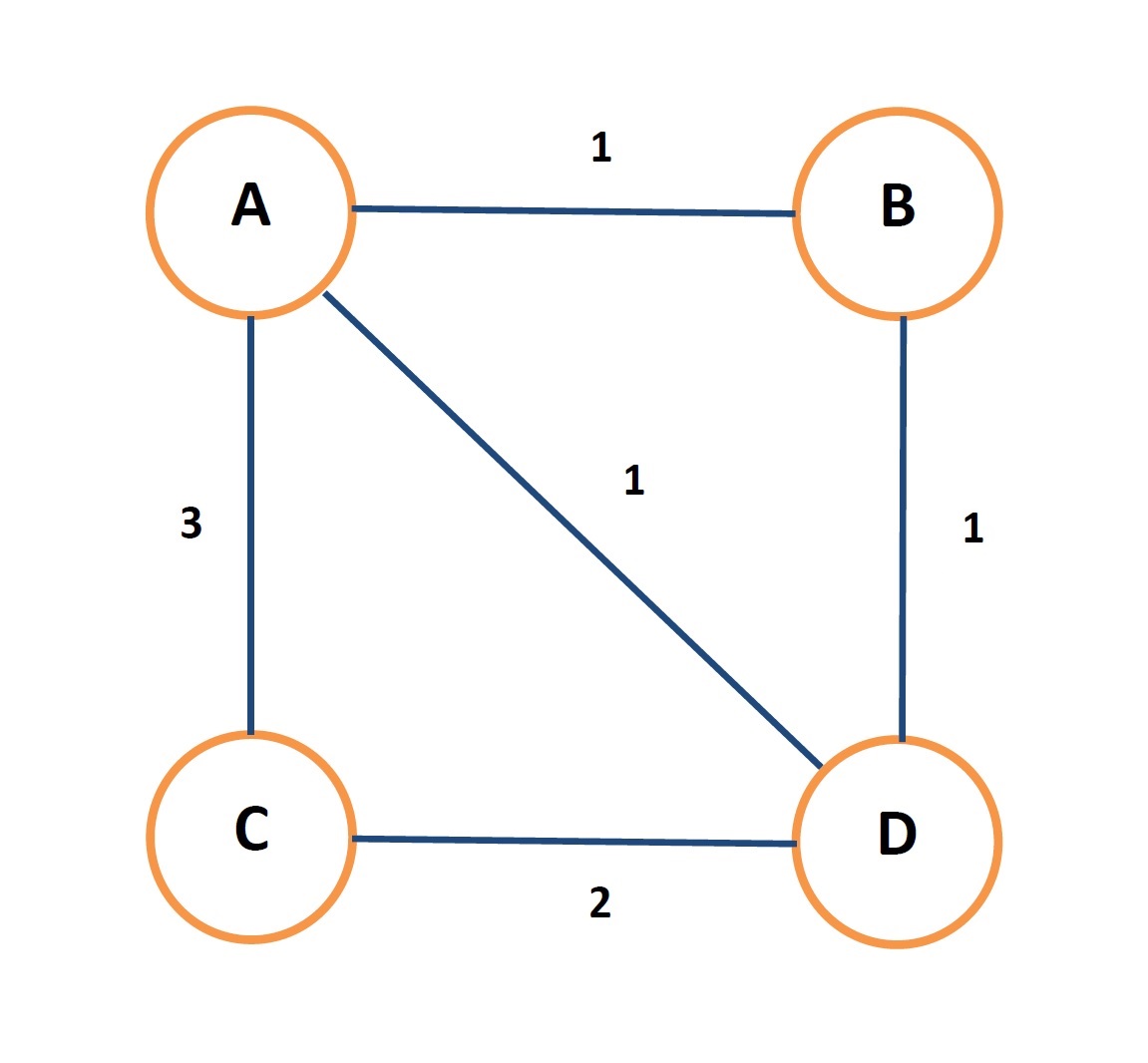}
\end{center}
 \[ A: 000 \quad  B: 001 \quad  C: 111 \quad     D: 100\]
Since we have an edge with a weight of $3$, we need at least $3$ bits. Therefore, for this graph, we have $c(G,\mathbf{w})=3$.
\end{example}
\begin{observation}\label{induce}
Let $(H,\mathbf{w^\prime})$ be an induced subgraph of $(G,\mathbf{w})$, then $c(H,\mathbf{w^\prime})\leq c(G,\mathbf{w})$.
\end{observation}
Our problem has an equivalent formulation in terms of edge partitioning of weighted graphs, as follows.
\begin{problem}
    For any weight-minimal graph $G$, $c(G,\mathbf{w})$ is equal to the minimum number of simple bipartite graphs that partition all edges of distance multigraph of $G$.
\end{problem}
\section{Main Results}\label{sec3}
In this section, we present the main results of the paper. To better organize the presentations, we split the results into three parts. First, we drive upper and lower bounds for $c(G,\mathbf{w})$ for the general graphs. This will help to find the exact value for this parameter for certain graphs. Next, we investigate the linear programming version of our problem. Finally, we study the parameter $c(G,\mathbf{w})$ for the Cartesian product of graphs.
\subsection{Results about $c(G,\mathbf{w})$}
In this subsection, we derive upper and lower bounds on $c(G,\mathbf{W})$ in terms of other graph theoretic parameters and also the exact value of it for some family of graphs. Later in Subsection \ref{cartpro}, we utilize the results of this subsection to derive an exact value of $c$ for other graph families.
\subsubsection{Lower and Upper Bounds on $c(G,\mathbf{w})$}
In this part, as a warm-up, we present one upper and two lower bounds on $c(G,\mathbf{w})$ for an arbitrary connected weighted graph $(G,\mathbf{w})$. In the subsequent sections, we derive tighter bounds. 
\begin{lemma}\label{threepart}
For every connected weighted graph $(G,\mathbf{w})$, we have:
\begin{enumerate}
    \item $c(G,\mathbf{w}) \geq \di_\mathbf{w}(G)$
    \item $c(G,\mathbf{w}) \geq \lceil \log |V|\rceil$
    \item $c(G,\mathbf{w}) \leq h_p(G_\mathbf{w}, \mathbf{w})$
    \item $c(G,\mathbf{w}) \leq t.(n-1)$ in which $t = \min _{T} \max_{e\in T} w(e)$. The minimum is taken over all spanning trees $T$ of $G$. 
\end{enumerate}
\end{lemma}
\begin{proof}
By considering the distance of the addresses of two nodes at $\di_\mathbf{w}(G)$, we can conclude part 1. \\
The part 2 is also trivial since different vertices must receive distinct addresses. Note that for a complete graph with constant weights $\mathbf{1}$, the bound holds with equality.
\\
For part 3, consider the lowest-weight Hamilton path in the graph. For this path, the following simple algorithm gives valid addresses of length $h_p(G_\mathbf{w}, \mathbf{w})$.
For the first vertex of path $v_1$, we assign an all-zero address of length $h_p(G_\mathbf{w}, \mathbf{w})$. Next, for the second vertex in the path, which is connected to $v_1$ with the edge of weight $w_1$, we assign an address with the same length of $h_p(G_\mathbf{w}, \mathbf{w})$ with exactly $w_1$ ones at the end of the code and so forth as below:
\begin{center}
$ \Bigl\{ \underbrace{0\dots0}_{h_p(G_\mathbf{w}, \mathbf{w})\rm\ times}, \quad  {0\dots0}\underbrace{1\dots1}_{h_p(G_\mathbf{w}, \mathbf{w})- w_1\rm\ times},\dots, \underbrace{1\dots1}_{h_p(G_\mathbf{w}, \mathbf{w})\rm\ times}  \Bigl\}$
\end{center}
Note that for this address assignment, for any two nodes $u,v$, there exists at least one path (namely the Hamilton path which we started with) such that the Hamming distance of the assigned addresses matches the distance of $u,v$ on that path. 
\\
Finally, for the last part, we first take $T$ to be a spanning tree whose maximum weight edge has the least weight among all spanning trees of $G$. By removing all the edges of $G$ outside $T$, we can only increase the value $c$; i.e. $c(G,\mathbf{w})\leq c(T,\mathbf{w})$. The assertion immediately follows by the fact that for trees, the binary addressing of the unweighted trees is no more than the addressing of it which is known to be equal to $n-1$ (See \cite{grah}). The extra $t$ factor simply compensates for the fact that the result of addressing number of trees only works for unweighted trees but here, the maximum weight of $T$ is $t$.
\end{proof}
\begin{corollary}
For the weighted path graph $(P,\mathbf{w})$, we have $c(P,\mathbf{w})=\di_\mathbf{w}(P)$.
\end{corollary}
\begin{proof}
The first lower bound in the above theorem matches the upper bound given in the mentioned algorithm. Thus, the corollary follows trivially.
\end{proof}
\begin{proposition}
    $c(G,\mathbf{1})=\log |V(G)|=m$ if and only if hypercube $Q_m$ is a subgraph of $G$.
\end{proposition}
\begin{proof}
    If $Q_m \subseteq G$, then $m\le c(G,\mathbf{1}) \le \log |V(G)|=m$. Also, when $c(G,\mathbf{1})$ is exactly $\log |V(G)|=m$, it means that all binary codes with length $m$ are used. As we should have the graph distance of any pair of vertices less than their Hamming distance, it implies that $Q_m$ is an induced subgraph of $G$.
\end{proof}
In the next example, we find the $c$ for the cocktail party graph which is a complete $m$-partite
graphs, where each part has exactly $2$
vertices. This graph is denoted by $K_{m\times2}$. This graph is equivalent to $K_{2m}$ deleting a perfect matching.
\begin{example}
    Let $(K_{m\times2}, \mathbf{1})$ be a simple cocktail party graph. Then we have $c(K_{m\times2}, \mathbf{1})=\lceil \log_2{2m}\rceil$.
\end{example}
\begin{example}
    For a simple cocktail party graph $(K_{m\times2}, \mathbf{1})$, the distance multigraph will be a $K_{2m}$ and a perfect matching. Its edges can be partitioned with exactly $\lceil \log_2{2m}\rceil$ simple bipartite graphs. 
\end{example}
\begin{remark}
    The addressing number for cocktail party graphs is still unknown. One can see the upper and lower bounds for that in \cite{zaks} and \cite{hoffman}.
\end{remark}
\subsubsection{Cycle Graphs}
In this part, we compute the exact value of $c(C_n, \mathbf{w})$ for every cycle $C_n$ and any arbitrary weight function $\mathbf{w}$. In Corollary~\ref{hamcyc} we use this result to obtain a tighter upper bound for $c(G,\mathbf{w})$ for any arbitrary graph $G$. We start by taking $\mathbf{w}$ as the constant weight $\mathbf{1}$.
\begin{lemma}\label{cycle}
For any integer $n\geq3$ we have $c(C_{n},\mathbf{1})=\lceil\frac{n}{2}\rceil$.
\end{lemma}
\begin{proof}
The lower bound follows from part 1 of Lemma \ref{threepart}. For the upper bound, we proceed as follows: We start from an arbitrary vertex and assign the all-zero binary address of length $\lceil\frac{n}{2}\rceil$ to it. Then, we traverse the cycle in one direction, and every time we reach a vertex, we assign an address by changing the leftmost $0$ bits of the previous address to $1$. At some point, we will assign all-one address to some vertex. From that point on, we change the leftmost $1$ entries of the previous address to $0$. For instance, for $C_6$, we have the following addresses assigned to its nodes.
\begin{center}
    \{000, 001, 011, 111, 110, 100\}
\end{center}
This construction guarantees that $c(C_{n},\mathbf{1})\leq \lceil\frac{n}{2}\rceil$. This bound matches the lower bound, and the result follows immediately.
\end{proof}
When the weight function is arbitrary, finding the exact value of $c(C_n,\mathbf{w})$ is more tricky. Besides the bounds in Lemma \ref{threepart}, we have the following upper bounds for $c(C_n,\mathbf{w})$.
\begin{lemma}
For any positive integer $n$ and weight function $\mathbf{w}$, we have:
\begin{enumerate}
    \item $c(C_n, \mathbf{w}) \leq \min_{u \in V(G)} \max_{u \neq v} d(u,v)$,
    \item $c(C_n, \mathbf{w}) \leq (\sum_{i=1}^{n} w_i) - \max_{i} w_i$.
\end{enumerate}
\end{lemma}
\begin{proof}
For 1, suppose that $ \min_{u \in V(G)} \max_{u \neq v} d(u,v)=k$ and this value is attained at two vertices $u^\prime$ and $v^\prime$. Therefore, there are two weighted paths along the cycle from $u^\prime$ to $v^\prime$. One with weight $k$ and another with weight less than or equal to $k$. We assign all-zero address with length $k$ to $u^\prime$ and all-one address with the same length to $v^\prime$. Consider an arbitrary vertex such as $z^\prime$ along the path from $u^\prime$ to $v^\prime$ with a larger weight. The minimum weighted path from $u^\prime$ to $z^\prime$ is along this larger weight path. The reason is that $k$ is the minimum of all maximum weighted paths between any pair of vertices. Moreover, the minimizer pair is $(u^\prime, v^\prime)$. Also, we can conclude that the minimum weighted path from $z^\prime$ to $v^\prime$ is along this larger weight path from $u^\prime$ to $v^\prime$. Therefore, the addresses with length $k$, starting from all-zero address and adding ones from both sides to it, are the binary addresses for $(C_n,\mathbf{w})$, and we have: $ c(C_n, \mathbf{w}) \leq \min_{u \in V(G)} \max_{u \neq v} d(u,v)=k$
\\
The second part of the lemma follows from part 3 of Lemma \ref{threepart}.
\end{proof}
Next, we find the exact value of $c(C_n,\mathbf{w})$. We do so, first, for $n=3$. Then, we use this case to extend the result to arbitrary values of $n$.
\begin{lemma}\label{k3}
For graph $C_3$, when the weights on edges are $\{a, b, c\}$ such that they satisfy the triangle inequality, $c(C_3, \mathbf{w})=\lceil \frac{a+b+c}{2} \rceil$.
\end{lemma}
\begin{proof}
Without loss of generality, suppose that $a\leq b \leq c$. Let $l=c(C_3,\mathbf{w})$ and $c_1, c_2, c_3$ be binary addresses of length $l$ for the vertices $u, v, z$ of $C_3$, respectively. Therefore, we have:
   \[ c \leq d(c_1, c_2), 
   \quad b \leq d(c_1, c_3), \quad
    a \leq d(c_2, c_3)\]
By summing up these inequalities, we obtain that $a+b+c \leq d(c_1, c_2)+d(c_2, c_3)+d(c_1, c_3) \leq 2l$. The last inequality is due to the fact that each coordinate of the addresses contributes at most $2$ to the summation $d(c_1, c_2)+d(c_2, c_3)+d(c_1, c_3)$.\\
On the other hand, $3$ codes with the length $k$ can be:
\begin{center}
$\underbrace{0\dots0}_{k \rm\ times}, \quad \underbrace{1\dots1}_{c \rm\ times}0\dots0, \quad 0\dots0\underbrace{1\dots1}_{b \rm\ times}$
\end{center}
The equality in the theorem follows immediately.
\end{proof}
\begin{theorem}\label{cyc}
For every weighted cycle $(C_n, \mathbf{w})$ in which the weight function $\mathbf{w}$ satisfies the triangle inequality, we have $c(C_n, \mathbf{w}) = \lceil \frac{\sum_{i=1}^{n} w_i}{2} \rceil$.
\end{theorem}
\begin{proof}
First of all, the weight-minimal graph satisfies the triangle inequality. We give lower and upper bounds for $c(C_n, \mathbf{w})$ as follows:
\begin{enumerate}
\item
For any pair of distinct vertices $u,v$ on a cycle $C_n$, there are two $(u,v)$-paths on $C_n$. Let $f(u,v)$ and $F(u,v)$ be the lengths of the shorter and the longer $(u,v)$-paths, respectively. Clearly, $f(u, v) + F(u, v) = \sum_{i=1}^{n} w_i $.
\\
Let $u,v$ be two vertices on the cycle such that they maximize $f(u,v)$. This pair is a minimizer for $F(u,v)$.
\\
The cycle $C_n$ has two $(u,v)$-paths, which we call $C_1$ and $C_2$. Assume that $C_2$ is the path of larger weight. We claim that $C_2$ contains at least one more vertex. If not, then $C_2$ is a single edge with a weight more than the $(u,v)$-path, $C_1$ which violates the triangle inequality. Let $z$ be the internal vertex of $C_2$. Consider the vertices $u,z$. Let $C_3$ be the $(u,z)$-path on $C_n$ which passes through the vertex $v$. Also, let $C_4$ be the other $(u,z)$-path on the cycle. We claim that the weighted distance between $u$ and $z$ is equal to the weight of $C_4$. If this is not the case, we must have $w(C_3) < w(C_4)$. On the other hand, $C_3$ includes $C_1$. Therefore, $w(C_1) \leq w(C_3)$. So, $f(u,z) > f(u,v)$. This contradicts the fact that $u,v$ maximizes $f(u,v)$.
\\
Similarly, if we consider the pair $v,z$, we conclude that their weighted distance is equal to the weight of the $(v,z)$-path that does not path through $u$.
\\
Now, consider, a $3$-cycle, $u^\prime v^\prime z^\prime$ such that $d(u^\prime, v^\prime)=d_{C_n} (u,v)$, $d(u^\prime, z^\prime)=d_{C_n} (u,z)$, and $d(v^\prime, z^\prime)=d_{C_n} (v,z)$. Clearly, each binary addressing of $C$ includes a valid addressing for this $3$-cycle. Since we already know the answer for $3$-cycles by Lemma \ref{k3}, we conclude that we need at least $\lceil \frac{\sum_{i=1}^{n} w_i}{2} \rceil$ bits for these $3$ vertices on the cycle.
\item
Here, we propose the following assignment structure: 
\\
Instead of each edge with a weight $w_i >1$, we put $w_i -1$ vertices on that edge. Then, we put a weight of $1$ on all those $w_i$ edges.
\\
The result will be a simple cycle with $n+ \sum_{i=1}^{n} (w_i -1)$ vertices and weights of all $1$. For this simple cycle by Lemma \ref{cycle}, we need $\lceil \frac{\sum_{i=1}^{n} w_i}{2} \rceil$ bits. The fact that this address assignment is valid is straightforward.
\end{enumerate}
These two parts imply the result of the theorem.
\end{proof}
\begin{corollary}\label{hamcyc}
For every connected weighted graph $c(G,\mathbf{w})$, we have $c(G,\mathbf{w}) \leq \lceil \frac{h_c(G_\mathbf{w}, \mathbf{w})}{2} \rceil$.
\end{corollary}
In the next example, our cases are weighted graphs with $4$ vertices.
\begin{example}\label{kfour}
Let $(G,\mathbf{w})$ be a connected weighted graph with $4$ vertices. Then, $(G,\mathbf{w})= \max_{i=1}^{4} \lceil \frac{T_i}{2} \rceil$, where $T_i$ is the weight of each $C_3$, subgraph of $G$ by removing the $i$-th vertex from $G$.
\end{example}
\begin{proof}
Let $u, v, z$ be a maximum weight triangle subgraph of $G_\mathbf{w}$ such that the weights of edges $uv$, $vz$, $uz$ in $G_\mathbf{w}$ are $c, a, b$, respectively. \\
We consider two cases:
\begin{enumerate}
\item
If $a \geq uw, b \geq vw, c \geq zw$ then, in this scenario, any addresses for the fourth vertex $w$, in the graph with weights $a, b, c, uw=a, vw=b, zw=c$ works for our graph with the given weights. It is easy to check that for this revised graph with mentioned weights, addresses with length $\lceil \frac{a+b+c}{2} \rceil$ will work.
\item
If $zw= c+ \gamma$, since we have the maximum sum of weights on $\vartriangle uvz$, and our triangle inequality condition holds, we have $uw \leq a- \gamma$ and $vw \leq b -\gamma$. Thus, we assign the address $c(u)+c(v)+c(z)+ \Bar{\gamma}$ to the fourth vertex $w$ (Here $c(u)$ is the address on the vertex $u$, and $\Bar{\gamma}$ is a binary vector with ones in those positions that $c(u)$ and $c(v)$ have ones in common). Therefore, in this case, the length of addresses will be $\lceil \frac{a+b+c}{2} \rceil$.
\end{enumerate}
\end{proof}
\subsection{Results about $\lambda$-BSEP}\label{lambdasec}
Consider a connected weighted graph $(G,\mathbf{w})$ and a positive integer number $\lambda$. By a $\lambda$-BSEP of $(G,\mathbf{w})$, we mean the problem of labeling the vertices of $G$ by the shortest possible length binary strings (i.e. a function $\phi: V(G) \rightarrow \{0,1\}^l$) such that for every pair of $u,v \in V(G)$, $\lambda  d_G(u,v) \leq d_H\big(\phi(u),\phi(v)\big)$.
\\
The minimum value of $l$ for the parameter $\lambda$ is denoted by $c_{\lambda}(G,\mathbf{w})$.
Observe that $c_{1}(G,\mathbf{w})$ is precisely $c(G,\mathbf{w})$ and in general, $c_{\lambda}(G,\mathbf{w})=c(G,\lambda .\mathbf{w})$. The following lemma states the subadditivity behavior of $c_{\lambda}$ when we fix the weighted graph but we let $\lambda$ vary. 
\begin{lemma}
Let $\lambda_1$, $\lambda_2$ be two positive integers. Then, we have $c_{\lambda_{1}+\lambda_{2}}(G,\mathbf{w}) \leq c_{\lambda_{1}}(G,\mathbf{w}) + c_{\lambda_{2}}(G,\mathbf{w})$.
\end{lemma}
\begin{proof}
Let $\phi_i: V(G) \rightarrow \{0,1\}^{c_{\lambda_{i}}(G,\mathbf{w})}$ be labeling of $\lambda$-BSEP for $(G,\mathbf{w})$, when $i=1,2$. Now, the concatenation $\phi_1$ and $\phi_2$ is indeed a $\big(\lambda_{1}+\lambda_{2}\big)$-labeling for $(G,\mathbf{w})$.
\end{proof}
This property of $c_{\lambda}$ is called ``sub-additivity'', which has interesting consequences. The next lemma is one such property:
\begin{lemma}\label{candb}
For every $\lambda > 0$, and connected weighted graph $(G,\mathbf{w})$, we have $c(G,\mathbf{w}) \geq \frac{c_{\lambda}(G,\mathbf{w})}{\lambda}$.
\end{lemma}
\begin{proof}
Using the sub-additivity property of $c_{\lambda}$ and by induction on $\lambda$, we have $c_{\lambda}(G,\mathbf{w}) \leq \lambda c_{1}(G,\mathbf{w}) = \lambda c(G,\mathbf{w})$. 
\end{proof}
A corollary of the previous two lemmas is the following one:
\begin{corollary}\label{lanw}
For every connected weighted graph $(G,\mathbf{w})$ and every positive integer $\lambda$, we have $c(G,\mathbf{w}) \geq \frac{c(G,\lambda .\mathbf{w})}{\lambda}$.
\end{corollary}
Inwards, Corollary \ref{lanw} says that the $c_{\lambda}$, normalized by $\lambda$ is a lower bound on $c(G,\mathbf{w})$. This is analogous to other graph theoretical parameters and their fractional counterparts such as a fractional chromatic number. As we will see in the next section, this concept can be best described using integer programming formulation and its linear relaxation. 
\\ 
Another consequence of the sub-additivity of $c_{\lambda}$ is the asymptotic behaviour of $\frac{c_{\lambda}(G,\mathbf{w})}{\lambda}$. For this, we need to use the following results about sub-additive sequences.
\begin{lemma}[Fekete \cite{feke}]
Let $\{a_n\}$ be a sequence of non-negative real numbers with the sub-additive property, $a_i + a_j \geq a_{i+j}$ for all $i,j \geq 1$. Then, $\lim_{n \rightarrow \infty} \frac{a_n}{n}$ exists and equals to $\inf_{n\geq1} \frac{a_n}{n}$.
\end{lemma}
\begin{corollary}
For any connected weighted graph $(G,\mathbf{w})$, the $\lim_{\lambda \rightarrow \infty}\frac{c_{\lambda}(G,\mathbf{w})}{\lambda}$ exists.
\end{corollary}
\begin{proof}
The proof is immediate since all the assumptions of Fekete's lemma are satisfied.
\end{proof}
Denote this limit by $\beta(G, \mathbf{w})$. That is $\lim_{\lambda \rightarrow \infty}\frac{c_{\lambda}(G,\mathbf{w})}{\lambda}:=\beta(G, \mathbf{w})$.
\\
Note that by Lemma \ref{candb}, we have $c(G,\mathbf{w}) \geq \frac{c_{\lambda}(G,\mathbf{w})}{\lambda}$. By taking different values of $\lambda$, we get different lower bounds for $c(G,\mathbf{w})$. Therefore, one might ask for $\lambda$, which $\frac{c_{\lambda}(G,\mathbf{w})}{\lambda}$ is the minimum.
\begin{example}\label{k3one}
Consider the cycle graph $C_3$ with all $\mathbf{1}$ as weights. The $c_{2}$ for this graph is $3$. Therefore, we have $\frac{c_{2}(C_3,\mathbf{1})}{2}=\frac{3}{2}$. If we consider $\lambda=3$ by Lemma \ref{k3}, we get $c_{3}(C_3,\mathbf{1})=5$. Then, we have $\frac{c_{3}(C_3,\mathbf{1})}{3}=\frac{5}{3}$.
\\
On the other hand, by Lemma \ref{k3} and \ref{candb}, we have $c_{\lambda}(C_3,\mathbf{1})= \lceil \frac{3 \lambda}{2}\rceil \geq \frac{3 \lambda}{2}$. Therefore, $\inf_{n\geq1}\frac{c_{\lambda}(C_3,\mathbf{1})}{\lambda}=\frac{3}{2}$.
\end{example}
For general graphs, in the next subsection, we will come back to this question.
\\
The next theorem is a stronger version of Lemma \ref{threepart}.
\begin{theorem}
For every connected weighted graph $(G,\mathbf{w})$:
\begin{enumerate}
    \item $c_{\lambda}(G,\mathbf{w}) \geq \lambda \di_\mathbf{w} (G)$ for all positive integer $\lambda$.
    \item $c_{\lambda}(G,\mathbf{w}) \leq 2 \lambda \di_\mathbf{w} (G)$ for some large enough $\lambda$.
\end{enumerate}
\end{theorem}
\begin{proof}
For 1, the reason based on Lemma \ref{threepart}, is as follows:
\begin{align*}
\lim_{\lambda \rightarrow \infty} \frac{c_{\lambda}(G,\mathbf{w})}{\lambda}\geq\lim_{\lambda \rightarrow \infty} \frac{ \max_{u \neq v} d_{G_{\lambda .\mathbf{w}}}(u,v)}{\lambda} 
=\max_{u \neq v} d_{G_\mathbf{w}}(u,v)=\di_\mathbf{w}(G)
\end{align*}
For 2, the main tool to achieve the upper bound for $\beta(G, \mathbf{w})$ is Hadamard codes. A Hadamard code of length $m$ is a set of $m+1$ vectors in $\{0,1\}^{m+1}$ such that the Hamming distance between any two of them is equal to $\frac{m+1}{2}$.
\\
It is known that for any positive integer $k$, a Hadamard code with $m=2^k -1$ exists. (see \cite{hamm}). Now, let $m=2^k -1$ be an integer such that $m \geq \max \big\{n, 2 \lambda \di_\mathbf{w} (G)\big\}$. Let $\mathcal{H}$ be a Hadamard code of length $m$.
\\
By the choice of $m$, $\mathcal{H}$ has at least $n$ elements. Pick any subset of $\mathcal{H}$ of size $n$ and call its elements $x_1, x_2, \dots, x_n$. Define $\phi: V(G) \rightarrow \{0,1\}^{m}$, $\phi(v_i):=x_i$. 
\\
For any two vertices $v_1, v_2$ we have $d_H(x_i,x_j) = d_H\big(\phi(u),\phi(v)\big)= \frac{m+1}{2} \geq \lambda \di_\mathbf{w} (G) $. 
Thus, $c_{\lambda}(G,\mathbf{w}) \leq m = \max \big\{n, 2 \lambda \di_\mathbf{w} (G)\big\}$. Therefore, $\beta(G, \mathbf{w}) \leq \frac{c_{\lambda}(G,\mathbf{w})}{\lambda} \leq \frac{\max \{n, 2 \lambda \di_\mathbf{w} (G)\}}{\lambda} = 2 \di_\mathbf{w} (G)$.
\\
The last inequality holds when $\lambda$ is large enough.
\end{proof}
Notice that for a small value of $\lambda$, the maximum of $n$ and $2 \lambda \di_\mathbf{w} (G)$ might be $n$. However, by increasing $\lambda$, the term $n$ becomes irrelevant. This is why in part 2 of Lemma \ref{threepart} we have the lower bound $\lceil \log n \rceil$ which depends on the size of the graph while for the fractional version (i.e. normalized $\lambda$-BSEP), we do not have the analogous lower bound.
\begin{corollary}\label{beta}
For every connected weighted graph $(G,\mathbf{w})$, we have:
\begin{center}
$\di_\mathbf{w} (G) \leq \beta(G, \mathbf{w}) \leq 2 \di_\mathbf{w} (G) $
\end{center}
\end{corollary}
\begin{example}
As we saw in the proof of part 2 of Lemma \ref{threepart}, $c(K_n,\mathbf{1})=\lceil \log n \rceil$ while $\beta(K_n, \mathbf{1}) \leq 2 \di_\mathbf{w} (K_n)=2$.
\end{example}
\begin{remark}\label{betacycles}
    For any weighted cycle $(C_n,\mathbf{w})$, we have $\beta(C_n, \mathbf{w})=\frac{\sum_{i=1}^{n} w_i}{2} $ and when the sum of weights is an even number, we have the equality of $\beta(C_n,\mathbf{w})$ and $c(C_n,\mathbf{w})$.
\end{remark}
\subsubsection{Integer Programming Formulation and Its  Linear Relaxation}\label{lp}
 We aim to express our problem as an integer program in this section. Suppose that there exist valid binary addresses for the weighted graph $(G,\mathbf{w})$ with $n$ vertices of length $l$. Therefore, we have a $n \times l$ binary matrix $M$, which we aim to minimize the $l$. Now, take any subset $A \subseteq \{1, 2, \dots, n\} = [n]$. Also, take the $i$-th column of the matrix $M$, as $c_i$. Any $c_i$ has some $1$'s, we put those numbers of coordinates in the set $C_i$. Next, we define $S_A$ for each subset $A \subseteq [n]$ as below:
 \begin{center}
 $S_A:=|\{i | C_i=A\}|$, where $|.|$ is the cardinal of a set.
 \end{center}
We can observe that for any two distinct sets $A$ and $B$ and a fixed $i$, the intersection of $\{i | C_i=A\}$ and $\{i | C_i=B\}$ is empty. This is because each column can only have one set of coordinates for the $1$'s, and this set cannot be equal to both $A$ and $B$ unless they are the same sets.
\\
Now, let $v_i$ and $v_j$ be two distinct vertices of $G$. By definition, the number of positions $k$, in which one of the addresses of $v_i$, $v_j$ is $1$, and the other one is $0$ must be at least $d_G(v_i,v_j)=w_{ij}$. In mathematical notation, we have:
\begin{center}
    $\forall \hspace{0.5em} 1\leq i < j \leq n, \quad \sum_{i\not\in A, j \in A} S_A + \sum_{j\not\in B, i \in B} S_B \geq w_{ij} $
\end{center}
Also, clearly, $S_A \in \mathbb{Z}$.
\\
Furthermore, the summation of all $S_A$'s is equal to the length of the addresses since each coordinate has a contribution $1$ to exactly one $S_A$.
\\
Conversely, if there exists an integer solution for the following integer program $IP_1$, one can immediately construct valid addresses for $G$.
$$  \min \quad  l = \sum_{A\subseteq [n]} S_A $$
\[
\mathrm{s.t.} \quad \sum_{i\notin A, j \in A} S_A + \sum_{j\notin B, i \in B} S_B \ge w_{ij}, \quad \forall \hspace{0.5em} 1 \le i < j \le n,\]
\\
\[ \quad S_A \in \mathbb{Z}_{\ge 0},  \quad \forall \hspace{0.5em} A \subseteq [n]. \] 

We may summarize the above discussion in the following theorem:
\begin{theorem}\label{linearc}
For every connected weighted graph $(G,\mathbf{w})$, $c(G,\mathbf{w})=OPT_{I_1}$, where $OPT_{I_1}$ is the optimal solution of the integer program $IP_{1}$.
\end{theorem}
Analogously, we define integer and linear programs for $\lambda$-scale binary addressing problems. Namely, let $IP_{\lambda}$ and $LP_{\lambda}$ be the following integer and linear programs indexed by the positive integer $\lambda$.

        $$ \min \quad  l_{\lambda} = \sum_{A\subseteq [n]} S_A$$
        \\
        \[ \mathrm{s.t.} \quad \sum_{i\notin A, j \in A} S_A + \sum_{j\notin B, i \in B} S_B \ge \lambda w_{ij}, \quad \forall \hspace{0.5em} 1 \le i < j \le n,\]
        \\
         \[S_A \in \mathbb{Z}_{\ge 0}, \quad \forall\hspace{0.5em} A \subseteq [n]. \]

Observe that Theorem \ref{linearc} implies that $OPT_{I_\lambda}$ is equal to $c(G, \lambda .\mathbf{w} )$. where $OPT_{I_\lambda}$ is the optimal value of $IP_{\lambda}$. 
Let $\mathcal{P}_\lambda$ be the feasible region of $LP_{\lambda}$. Thus, $OPT_{I_\lambda}$ is a point of $\mathcal{P}_\lambda$, which minimizes the objective function $\underset{A \subseteq [n]}{\sum S_A}$.
\\
Similarly, $IP_{\lambda}$ is a lattice point (i.e. a point with integral coordinates) of $\mathcal{P}_\lambda$ with minimum sum of coordinates. 
\\
We know from linear programming that the minimum of a linear function over a polyhedron if it exists, is attained at a corner point. Hence, if $\mathcal{P}_\lambda$ is an integral polyhedron (i.e. a polyhedron with all the corner points to be integral), then $OPT_{I_\lambda}=OPT_{L_\lambda}$.
\\
For arbitrary integer weights, the corner points of $\mathcal{P}_1$ are not necessarily integral. However, since the corner points of a polyhedron defined by a set of linear constraints with rational coefficients are always rational (see \cite{mat}), the corner points of $\mathcal{P}_1$ are rational. Let $\mu$ be the common denominator of all the coordinates of all the corners of $\mathcal{P}_1$. Using the next lemma, we will prove that the polyhedron $\mathcal{P}_{\mu}$ is integral.
\begin{lemma}
For every connected weighted graph $(G,\mathbf{w})$, and every positive integer $\lambda$ we have $\mathcal{P}_{\lambda} = \lambda \mathcal{P}_1$ where $\lambda \mathcal{P}_1= \{x| x = \lambda x^\prime \quad \text{for some} \quad x^\prime \in \mathcal{P}_1 \}$.
\end{lemma}
\begin{proof}
First, suppose that $x = \lambda x^\prime$ for some $x^\prime \in \mathcal{P}_1$.
Thus, $x^\prime$ satisfies all inequalities of the form
$\forall \hspace{0.5em} A: \sum_{i\not\in A, j \in A} x^\prime_A + \sum_{j\not\in A, i \in A} x^\prime_A \geq w_{ij} $.
\\
This implies that: $\forall \hspace{0.5em} A \subseteq [n], \forall \hspace{0.5em} i\neq j \in [n], \sum_{i\not\in A, j \in A} \lambda x^\prime_A + \sum_{j\not\in A, i \in A}  \lambda x^\prime_A \geq  \lambda w_{ij}$. Therefore, $\lambda x^\prime \in \lambda \mathcal{P}_1 $. Conversely, a similar argument shows that if $x \in \mathcal{P}_{\lambda}$, then $x = \lambda x^\prime$ for some $x^\prime \in \mathcal{P}_1$. This completes the proof.
\end{proof}
\begin{corollary}
For every connected weighted graph $(G,\mathbf{w})$ and every positive integer $\lambda$, $OPT_{I_\lambda} = \lambda OPT_{I_1}$.
\end{corollary}
\begin{proof}
Since the objective is a linear function, the assertion is a direct implication of the previous lemma.
\end{proof}
\begin{corollary}\label{integ}
For every connected weighted graph $(G,\mathbf{w})$, there exists an integer $\mu$ such that $\mathcal{P}_{\mu}$ is an integral polyhedron.
\end{corollary}
\begin{proof}
As we saw earlier, $\mathcal{P}_{1}$ has rational corners. Thus, if we consider $\mu$ as the least common multiple of all the coordinates of all the corners of $\mathcal{P}_{1}$, then $\mu \mathcal{P}_{1}= \mathcal{P}_{\mu}$ has integral corners.
\end{proof}
\begin{corollary}\label{mu}
For every connected weighted graph $(G,\mathbf{w})$, there exists a number $\mu$ such that $\beta(G, \mathbf{w})=\frac{c_{\mu}(G,\mathbf{w}) }{\mu}$.
\end{corollary}
\begin{proof}
Set $\mu$ to be the number defined in the previous corollary. Since $\mathcal{P}_{\mu}$ is integral by Corollary \ref{integ}, we know that $OPT_{L_{\mu}} =  OPT_{I_{\mu}} = c_{\mu}(G,\mathbf{w})$. On the other hand, $OPT_{L_{\mu}}= \mu \beta(G, \mathbf{w})$. Thus $\beta(G, \mathbf{w}) = \frac{c_{\mu}(G,\mathbf{w}) }{\mu} $. 
\end{proof}
What this corollary implies is that the best lower bound of the type $\frac{c_{\lambda}(G,\mathbf{w}) }{\lambda} $ for $c(G,\mathbf{w})$ is attained for some finite $\mu$. For instance, for $C_3$ with constant weight $\mathbf{1}$, we saw in Example \ref{k3one} that when $\lambda=2$ the minimum of $\frac{c_{\lambda}(C_3,\mathbf{1}) }{\lambda} $ is attained. 
\\
An interesting particular case is when $G$ is a complete graph of size $n$ and $\mathbf{w}=1$. Then, Corollary \ref{mu} implies that among all codes of size $n$, there exist two constants $m$ (length of code) and $r$ (the minimum distance of the code), such that $\frac{m}{r}$ is the minimum. Alternatively, $\frac{r}{m}$ is the maximum. In the field of error correction codes, the quantity $\frac{r}{m}$ is called the relative distance of the code.
\begin{example}
For $n=4$, we will try to find the largest relative distance among all the codes with $4$ codewords. As we saw, the maximum of $\frac{r}{m}$ is attained when $\frac{m}{r}$ is the minimum. The minimum of $\frac{m}{r}$ is precisely $\beta(K_4,\mathbf{1})$. We can now utilize Example \ref{kfour}, to get that $\beta(K_4,\mathbf{1})=\lim_{\lambda \rightarrow \infty} \frac{c_{\lambda}(K_4, \mathbf{1})}{\lambda} = \frac{\lceil \frac{3 \lambda}{2}\rceil}{\lambda} =\frac{3}{2}$.
\\
Hence, the best relative distance of a code of size $4$ is $\frac{3}{2}$, and is achieved by the following code:
\begin{center}
$\{(0,0,0), (1,1,0), (1,0,1), (0,1,1)\}$
\end{center}
\end{example}
\vspace{0.5cm}
Now, we consider the dual program of $LP_1$. The following linear maximization is the dual of $LP_1$ and we call it $LP^\prime_1$. \\
\begin{equation}
    \tag{$LP^\prime_1$}
    \begin{array}{ll}
         \max & \underset{1\leq i < j \leq n}{\sum w_{ij}z_{ij}} \\
         \mathrm{s.t.} & 
         \sum_{|\{i,j\} \cap A|=1 } z_{ij} \leq 1, \quad \forall \hspace{0.5em} A \subseteq [n],\\
         &  z_{ij} \geq 0, \quad \forall \hspace{0.5em} 1\leq i < j \leq n. 
    \end{array}
\end{equation}
\\
This problem has an interesting combinatorial interpretation. Think of $z_{ij}$'s as some non-negative numbers assigned to the edges of $G_\mathbf{w}$ (which is a complete graph). Let $A$ be some subset of $[n]$. This set corresponds to one of the constraints in the above program. The variables contributing to this constraint, (i.e. $\sum_{|\{i,j\} \cap A|=1 } z_{ij} \leq 1$) correspond to the edges of $G_\mathbf{w}$ that have exactly one endpoint in $A$. In other words, the constraints of $LP^\prime_1$ correspond to the cuts of the graph $G_\mathbf{w}$, and they require that for every cut $A$, the sum of the values $z_{ij}$'s corresponding to the edges in the cut, is no more than $1$. Thus, the optimal solution of $LP^\prime_1$ is exactly the solution to the following combinatorial problem.
\begin{problem}
    A complete weighted graph $H$ on $n$ vertices is given such that the weight of the edge $v_i v_j$ is equal to the positive integer $w_{ij}$. We want to assign positive numbers $z_{ij}$ to the edges of $H$ such that:
\begin{enumerate}
    \item For every cut $(A, A^c)$, the sum of $z_{ij}$'s on the edges of the cut is at most $1$.
    \item The quantity $\sum w_{ij}z_{ij}$ is maximized.
\end{enumerate}
\end{problem}
Clearly, every feasible solution of $LP^\prime_1$ is a lower bound for $OPT_{L_1}$, according to the weak duality theorem. In particular, let $z_{ij}:=\frac{1}{\lfloor \frac{n^2}{4}\rfloor}$. We claim that such $z_{ij}$'s form a feasible solution. This is simply because the size of the maximum cut of $H$ is $\lfloor \frac{n^2}{4}\rfloor$. Hence for every cut $(A, A^c)$, we have:
\begin{center}
    $\sum_{|\{i,j\} \cap A|=1 } z_{ij}= |A|(n-|A|) \leq \lfloor \frac{n^2}{4}\rfloor \times \frac{1}{\lfloor \frac{n^2}{4}\rfloor} =1$.
\end{center}
The objective function in this case is equal to $\sum w_{ij}z_{ij}=\frac{\sum w_{ij}}{\lfloor \frac{n^2}{4}\rfloor}$. Therefore, we get the following lower bound for $\beta(G, \mathbf{w})$.
\begin{theorem}\label{plot}
For every connected weighted graph $(G,\mathbf{w})$, if the weight of the edge $v_i v_j$ in $G_\mathbf{w}$ is $w_{ij}$, then $\beta(G, \mathbf{w}) \geq \frac{\sum w_{ij}}{\lfloor \frac{n^2}{4}\rfloor}$.
\end{theorem}
\begin{corollary}\label{boundbeta}
Under the assumptions of the previous theorem, for every subset $B\subseteq V(G)$, we have $\beta(G, \mathbf{w}) \geq \frac{\sum_{v_i, v_j \in B} w_{ij}}{\lfloor \frac{|B|^2}{4}\rfloor}$.
\end{corollary}
\begin{proof}
The corollary follows from the previous theorem and Observation \ref{induce}.
\end{proof}
In Appendix \ref{secA1}, we show that Theorem \ref{plot} is analogous to the Plotkin bound in the theory of error correction codes.
\begin{example}
Consider $K_4$ with weights of $\{w_{12}, w_{13}, w_{14}, w_{23}, w_{24}, w_{34}\}$ on its edges. The constraints of the Dual-LP for this graph are related to $4$ bipartite graphs in the form of $K_{1,3}$ and $3$ bipartite graphs in the form of $K_{2,2}$. In fact, for this case, the linear programming is as follows:
\begin{equation*}
    \begin{array}{ll}
         \max &w_{12}z_{12} + w_{13}z_{13} + w_{14}z_{14} + w_{23}z_{23} + w_{24}z_{24} + w_{34}z_{34}\\
         \mathrm{s.t.} & 
        z_{12} + z_{13} + z_{14} \leq 1, \quad
z_{12} + z_{23} + z_{24}  \leq 1, \\
&z_{13} + z_{23} + z_{34}  \leq 1, \quad 
z_{14} + z_{24} + z_{34}    \leq 1, \\
&z_{13} + z_{14} + z_{23} + z_{24}  \leq 1, \quad 
z_{12} + z_{14} + z_{23} + z_{34}      \leq 1, \\
&z_{12} + z_{13} + z_{24} + z_{34}      \leq 1, \quad
z_{ij} \geq 0, \quad \forall \hspace{0.5em} 1\leq i < j \leq 4.
    \end{array}
\end{equation*}
\end{example}

\subsubsection{Integrality Gap}
In a feasible linear minimization problem $M$, let $M_{int}$ be the minimum value at an integral feasible point. Also, denote the global minimum value by $M_{frac}$. The integrality gap of $M$ is defined as the ratio $IG:=\frac{M_{int}}{M_{frac}}$.
\\
In this part, we find bounds for the integrality gap of integer program $IP_{1}$ and its relaxed linear program $LP_{1}$. Since the problem is a minimization problem, the integrality gap $IG$ is always at least 1. The upper bound is provided in the following result.
\begin{proposition}
For a weighted graph with $n$ vertices, the integrality gap of integer program $IP_{1}$ and its relaxed linear program $LP_{1}$, is at most $\lceil \log n \rceil$.
\end{proposition}
\begin{proof}
Using Corollary \ref{beta}, we can infer that the lower bound of the relaxed minimum value of $LP_{1}$ is $\di_\mathbf{w} (G)$. Meanwhile, the upper bound of the integer minimum value of $IP_{1}$ is $\di_\mathbf{w} (G) \times \lceil \log n \rceil$. This is because we can assign distinct addresses to each vertex and repeat them $\di_\mathbf{w} (G)$ times. Hence, the length of binary addresses will be $\di_\mathbf{w} (G) \times \lceil \log n \rceil$.
\end{proof}
\begin{remark}
Note that for graphs with small edge weights but large diameter (for instance, for a simple path whose diameter is as big as $O(n)$ but the edge weights are all 1), the bound $\lceil \log(n)\rceil \times \di(G)$ is poor compared to the bound in part 4 of Lemma~\ref{threepart}.
\end{remark}
\subsection{Cartesian Product of Graphs}\label{cartpro}
Our goal in this section is to study the values of $c,\beta$ parameters under the Cartesian product operation. 
In the language of coding theory, one can concatenate two binary addresses of $(G_1,\mathbf{w}_1)$ and $(G_2,\mathbf{w}_2)$ to obtain one for $(G_1 \square G_2, \mathbf{w}_1 \square \mathbf{w}_2)$. This will give an upper bound for $c\big(G_1 \square G_2, \mathbf{w}_1 \square \mathbf{w}_2\big)$ ($\beta\big(G_1 \square G_2, \mathbf{w}_1 \square \mathbf{w}_2\big))$ in terms of $c(G_1,\mathbf{w}_1)$ and $c(G_2,\mathbf{w}_2)$ ($\beta(G_1,\mathbf{w}_1)$, $\beta(G_2,\mathbf{w}_2)$, respectively).
The next statement formalizes this simple observation.
\begin{observation}
For the Cartesian product of $k$ weighted graphs $(G_1, \mathbf{w}_1), \dots, (G_k, \mathbf{w}_k)$, we have: $$c\big(\square_{i=1}^{k} G_i, \square_{i=1}^{k} \mathbf{w}_i\big) \le \sum_{i=1}^{k}c(G_i, \mathbf{w}_i).$$
\end{observation}

What remains to answer is the question of whether these bounds are the exact answers.
\\
Although we do not know the answer, we present sufficient conditions under which $c\big(G_1 \square G_2, \mathbf{w}_1 \square \mathbf{w}_2\big)=c(G_1,\mathbf{w}_1)+c(G_2,\mathbf{w}_2)$, or $\beta\big(G_1 \square G_2, \mathbf{w}_1 \square \mathbf{w}_2\big)=\beta(G_1,\mathbf{w}_1)+\beta(G_2,\mathbf{w}_2)$. 
Here, we present some definitions.
\begin{definition}\label{def}
Let $f$ be a real-valued function on weighted graphs $(G,\mathbf{w})$ ($f(G)$ for short), then:
\begin{enumerate}[label=(\roman*)]
    \item We say a function $f$ is proper if:
    \begin{itemize}
        \item 
    For any $G_1,G_2$, $f(G_1 \square G_2) \ge f(G_1)+f(G_2)$,
    \item
    For all $G$, $f(G) \le c(G,\mathbf{w})$.
 \end{itemize}
\item 
We say a function $f$ is strongly proper if:
    \begin{itemize}
        \item 
    For any $G_1,G_2$, $f(G_1 \square G_2) \ge f(G_1)+f(G_2)$,
    \item
   For all $G$, $f(G) \le \beta(G,\mathbf{w})$.
 \end{itemize}
\item For a proper $f$, a graph is called $f$-proper when $f(G)=c(G,\mathbf{w})$.
\item For a strongly proper $f$, a graph is called $f$-strongly proper, when $f(G)=\beta(G,\mathbf{w})$.
\end{enumerate}
\end{definition}
Note that a strongly proper function is also a proper one since $\beta(G,\mathbf{w})\ge f(G,\mathbf{w})$ implies $c(G,\mathbf{w})\ge f(G,\mathbf{w})$. However, if $f$ is strongly proper and $(G,\mathbf{w})$ is $f$-strongly proper, then $(G,\mathbf{w})$ does not need to be $f$-proper. This is because $f(G,\mathbf{w})=\beta(G,\mathbf{w})$ does not necessarily implies that $f(G,\mathbf{w})=c(G,\mathbf{w})$. In fact, in most cases, it is not.
The following lemma provides some proper and strongly proper functions.
\begin{lemma}\label{types}
  Let $f$ be a real-valued function on weighted graphs. Then we have:
    \vspace{0.5em}
    \begin{enumerate}
        \item $f(G)=\lceil\log |V(G)|\rceil$ is proper. (When graph $G$ is unweighted.)
        \vspace{0.5em}
        \item 
        $f(G)=\di_\mathbf{w}(G)$ is strongly proper.
        \item $f(G)=\max_{\substack{H\subseteq G \\ |H|=2k}}$
        $\frac{\underset{e\in H}{\sum w_e}}{\frac{|H|^2}{4}}$ is strongly proper.
        \item 
    $f(G)=\max_{\substack{H\subseteq G \\ |H|=3}}$
    $\frac{\underset{e\in H}{\sum w_e}}{2}$ is strongly proper. 
    \end{enumerate}
\end{lemma}
\begin{proof}
$1.$ By definitions, the proof follows immediately.
\\
$2.$ For every graphs $G_1, G_2$, we have $\di_\mathbf{w}(G_1\square G_2)= \di_\mathbf{w}(G_1)+\di_\mathbf{w}(G_2)$. In addition to part 1 of Lemma \ref{threepart}, the result follows. 
\\
$3.$ From Corollary \ref{boundbeta}, we know this $f\le \beta(G,\mathbf{w})$ for each $G$. Also, consider this $f$ takes its maximum in $H_1 \subseteq G_1$ and $H_2\subseteq G_2$. Then, we have:
\[f(G_1 \square G_2) \ge \frac{\underset{e\in H_1 \square H_2}{\sum w_e}}{ \frac{|H_1|^2 |H_2|^2}{4}}=\frac{\underset{e\in H_2}{|H_1|^2\sum w_e}+\underset{e\in H_1}{|H_2|^2\sum w_e}}{\frac{|H_1|^2 |H_2|^2}{4}} =\frac{\underset{e\in H_2}{\sum w_e}}{\frac{|H_2|^2}{4}} +\frac{\underset{e\in H_1}{\sum w_e}}{\frac{|H_1|^2}{4}}=f(G_2)+f(G_1).\]
\\
$4.$ Similar to previous proof, $f\le \beta(G,\mathbf{w})$. Moreover, suppose this $f$ takes its maximum in triangles $T_1\subseteq G_1$ and $T_2\subseteq G_2$. Then:
\[f(G_1 \square G_2) \ge \frac{\underset{e\in T_1 \square T_2}{\sum w_e}}{2}\ge \frac{\underset{e\in T_2}{3\sum w_e}+\underset{e\in T_1}{3\sum w_e}+\underset{else}{\sum w_e}}{2} \ge \frac{\underset{e\in T_2}{\sum w_e}}{2} +\frac{\underset{e\in T_1}{\sum w_e}}{2}=f(G_2)+f(G_1).\]
\end{proof}
The following lemma is the main tool for the subsequent results in this section. It also clarifies the importance of Definition \ref{def} and Lemma \ref{types}.

\begin{lemma}\label{cartproper}
Let $f$ be a function on weighted graphs, then:
\begin{enumerate}
    \item 
  If $f$ is a proper function and $G_1, G_2$ are $f$-proper graphs, then $G_1\square G_2$ is $f$-proper graph too.
  \item 
  For the Cartesian product of graphs $(G_1,\mathbf{w}_1)$ and $(G_2,\mathbf{w}_2)$, we can conclude that,
$\beta (G_1 \square G_2, \mathbf{w}_1 \square \mathbf{w}_2) \leq \beta (G_1, \mathbf{w}_1) + \beta (G_2, \mathbf{w}_2)$.
\item
 If $f$ is a strongly proper function and $G_1, G_2$ are $f$-strongly proper graphs, then $G_1\square G_2$ is $f$-strongly proper graph too.
  \end{enumerate}
\end{lemma}
\begin{proof}
$1.$ By using Definition \ref{def}, we have
    $c(G_1)+c(G_2)\ge c(G_1 \square G_2)\ge f(G_1 \square G_2) \ge f(G_1)+f(G_2)=c(G_1)+c(G_2)$ $ \Rightarrow f(G_1 \square G_2)=c(G_1 \square G_2).$ Therefore, $G_1\square G_2$ is $f$-proper graph.
    \\
$2.$ Corollary \ref{mu}, and the concatenating addresses of vertices of graphs $G_1$, $G_2$ lead us to the result.
\\
$3.$ Similar to the first part of the proof, we can conclude that:\\
    $\beta(G_1)+\beta(G_2)\ge \beta(G_1 \square G_2)\ge f(G_1 \square G_2) \ge f(G_1)+f(G_2)=\beta(G_1)+\beta(G_2)$\\ $ \Rightarrow f(G_1 \square G_2)=\beta(G_1 \square G_2).$ Then, $G_1\square G_2$ is $f$-strongly proper graph.
\end{proof}
\begin{corollary}\label{compl}
For graph
$G = \square_{i=1}^{k} K_{n_i}$,
where $n_i= 2^{m_i}$,
we have
$c(G,\mathbf{1})=\sum_{i=1}^{k}c(K_{n_i},\mathbf{1})$.
\end{corollary}
\begin{proof}
    From part 2 of Lemma \ref{threepart} and part 1 of Lemma \ref{types}, we can conclude that $c(G, \mathbf{1})= \sum_{i=1}^{k} c(K_{n_i}, \mathbf{1})= \sum_{i=1}^{k} m_i$.
\end{proof}
\begin{remark}
Similar to the result of Corollary \ref{compl} about addressing number of the Cartesian product of graphs, i.e. $N\big(\square_{i=1}^{k}K_{n_i}\big)$ can be found in \cite{cioa}.
\end{remark}
\begin{corollary}
If $c(G_i,\mathbf{w}_i)=\di_\mathbf{w}(G_i)$ for all $1\leq i \leq k$, then we have $c\big(\square_{i=1}^{k} G_i, \square_{i=1}^{k} \mathbf{w}_i\big)=\sum_{i=1}^{k} c(G_i, \mathbf{w}_i)$.
\end{corollary}
\begin{proof}
To prove this statement, we part 2 of Lemma \ref{types} and the following chain of equalities and inequalities,
$\di_\mathbf{w}(\square_{i=1}^{k} G_i)=\sum_{i=1}^{k} \di_\mathbf{w}(G_i)\leq c\big(\square_{i=1}^{k} G_i, \square_{i=1}^{k} \mathbf{w}_i\big) \leq \sum_{i=1}^{k} c(G_i, \mathbf{w}_i)=\sum_{i=1}^{k} \di_\mathbf{w}(G_i)$.
\end{proof}
\begin{corollary}
For any number of weighted paths $(P_i,\mathbf{w}_i)$, we have $c\big(\square_{i=1}^{k} P_i, \square_{i=1}^{k} \mathbf{w}_i\big)=\sum_{i=1}^{k} c\big(P_i,\mathbf{w}_i\big)$.
\end{corollary}
\begin{proof}
 The proof follows from part 2 of Lemma \ref{types}.
\end{proof}

We conclude this subsection by the next two theorems and a corollary of them.

\begin{theorem}
Let $(C_n, \mathbf{w})$ and $(P_m, \mathbf{w^\prime})$ be weighted cycle and weighted path, respectively. Then, the parameter $c$ of $(C_n, \mathbf{w}) \square (P_m, \mathbf{w^\prime})$ is $c(C_n, \mathbf{w})+c(P_m, \mathbf{w^\prime})$.
\end{theorem}
\begin{proof}
We can apply part 4 of Lemma \ref{types} when the path $P_m$ has more than $3$ vertices. Consider the first, last, and a vertex in between these vertices. Then, the value of the strongly proper function in part 4 of Lemma \ref{types} is $\frac{2\sum_{i=1}^{m-1} w^\prime_i}{2}=\sum_{i=1}^{m-1} w^\prime_{i}$. Moreover, $(C_n, \mathbf{w})$ is $f$-proper graph too. Then, $C_n \square P_m$ is $f$-proper graph by Lemma \ref{cartproper}.
\end{proof}
\begin{theorem}
Let $(C_n, \mathbf{w})$ and $(C_m, \mathbf{w^\prime})$ be weighted cycles such that at least one of them is even (i.e. $\sum_{i=1}^{n} w_i=2k$ or $\sum_{i=1}^{m} \mathbf{w^\prime}_i=2k^\prime$). Then, we have $c\big((C_n, \mathbf{w}) \square (C_m,\mathbf{w^\prime})\big)=c(C_n, \mathbf{w})+c(C_m, \mathbf{w^\prime})$.
\end{theorem}
\begin{proof}
Consider $3$ vertices $u^\prime, v^\prime, z^\prime$ as the proof of Theorem \ref{cyc} on the first factor of $C_n$. Then, by considering these vertices; $u^\prime$ in the first factor of $C_n$, $v^\prime$ on some factor in the middle, and $z^\prime$ on the last factor in this graph Cartesian product, we get an induced triangle. The length of the binary addresses for this induced subgraph should be $\lceil\frac{\sum_{i=1}^{n} w_i +\sum_{i=1}^{m} w\prime_i}{2}\rceil$. Moreover, we have $c\big((C_n, \mathbf{w}) \square (C_m, \mathbf{w^\prime})\big)\le\lceil\frac{\sum_{i=1}^{n} w_i}{2}\rceil+\lceil\frac{\sum_{i=1}^{m} w\prime_i}{2}\rceil$. Due to the condition of this theorem, these upper and lower bounds for $c$ of the Cartesian graph match.
\end{proof}
\begin{corollary}
   Assume that each graph $(G_i,\mathbf{w}_i)$ has at most $4$ vertices or even weight sum, then we have $c\big(\square_{i=1}^{k} G_i, \square_{i=1}^{k} \mathbf{w}_i\big) = \sum_{i=1}^{k}c(G_i, \mathbf{w}_i)$.
\end{corollary}
\begin{theorem}
    For the Cartesian products of $k$ weighted cycles, we have
    $\beta\big(\square_{i=1}^{k} C_i, \square_{i=1}^{k} \mathbf{w}_i\big) = \sum_{i=1}^{k}\beta(C_i, \mathbf{w}_i)$.
\end{theorem}
\begin{proof}
    The proof is the result of Remark \ref{betacycles}, and part 4 of lemma \ref{types}.
\end{proof}
\section{Applications}
In this section, we show some ways that our problem can be applied to Lee metric codes which are very useful in DNA sequence storage.
\\
Recall the Lee metric codes, and the maximum number of Lee codes, denoted as $A_q^L (n, d)$. It is worth noting that any Lee code over the alphabet set $\mathbb{Z}_q$ corresponds to a set of vectors whose entries are vertices of a cycle graph with $q$ vertices. That is, we may identify the elements of $\mathbb{Z}_q$ with the vertices of a $q$-cycle such that the Lee distance of any two elements in $\mathbb{Z}_q$ is the same as the shortest distance between the corresponding vertices in the cycle. 

Imagine that two vectors of length $n$ over $\mathbb{Z}_q$, alternatively over the cycle $C_q$, of Lee distance $d$ is given. Suppose that a binary addressing of length $l$ is also provided. Thus, if we replace each coordinate of the initial vectors with the corresponding address, we obtain vectors that are of length $n.l$ and their pairwise Hamming distance is at least $d$. Replacing entries with their addresses will turn the vectors into longer ones over the binary alphabet while the distance (previously Lee distance, now, Hamming distance) is not reduced. 
The following relation exists between $A_2(n \lceil \frac{q}{2} \rceil, d)$ and $A_q^L (n, d)$.
\begin{theorem}\label{binarycycle}
Consider a simple cycle graph $C_q$, and $n,d\in\mathbb{N}$. We have $A_q^L (n, d) \leq A_2(n \lceil \frac{q}{2} \rceil, d)$.
\end{theorem}
\begin{proof}
Let us consider a Lee code with symbols that are the names of vertices, ${1, 2, \dots, q}$, and length $n$. Referring to lemma \ref{cycle}, it follows that the binary addresses of this graph have a length of $\lceil \frac{q}{2} \rceil$. Consequently, we can derive at least $A_q^L (n, d)$ binary codewords of length $n \lceil \frac{q}{2} \rceil$ and minimum distance $d$.
\end{proof}
In numerous papers, the lower and upper bounds of Lee metric codes have been investigated. By the previous theorem, we can improve the upper bound for $A_6^L (8, 14)$, which was less than or equal to $7$, using linear programming methods (refer to \cite{bound1}, \cite{bound2}, \cite{bound3}). We show that the upper bound is $6$. This is because there are exactly $6$ binary codewords with a length of $24$ and a minimum distance of $14$. Therefore, we have $A_6^L (8, 14) \leq A_2 (8\times 3, 14)=6$.
\begin{corollary}\label{weightedrel}
Consider a constant-weighted cycle graph $(C_q,\mathbf{w})$, and $n,d\in\mathbb{N}$. We have $A_q^L (n, d) \leq A_2(n \lceil \frac{wq}{2} \rceil, dw)$.
\end{corollary}
The best upper bound for Lee codes on $C_{17}$ with $n=4$ and $d=19$ is $11$ (See \cite{bound1}). By applying Theorem \ref{cyc} to $C_{17}$ with a constant weight of $2$, denoted as $c(C_{17},2)$, we obtain a length of $17$. Consequently, based on Corollary \ref{weightedrel}, we deduce that $A_{17}^L (4, 19) \leq A_2(68, 38)$. Furthermore, utilizing the well-known Plotkin bound, we find that $A_2(68, 38)\leq 2 \lfloor \frac{38}{38\times2 -68} \rfloor=8$. Hence, $A_{17}^L (4, 19) \leq 8$.
\\
Similarly, we improved other upper bounds for the size of Lee codes. Here, as compared to the results of \cite{bound1}, the following bounds in the table have been improved. To find all these improved bounds, we used Theorems \ref{binarycycle} and Corollary \ref{weightedrel}.
\vspace{5mm}
\\
\setlength{\arrayrulewidth}{0.3mm}
\setlength{\tabcolsep}{18pt}
\renewcommand{\arraystretch}{1.5}
\begin{tabular}{ |p{0.7cm}|p{0.7cm}|p{0.7cm}|p{1cm}|p{1.2cm}|p{1.2cm}|}
\hline
\multicolumn{6}{|c|} {New Upper Bounds Obtained by Our Method  for $A_{q}^L (n, d)$}\\
\hline
\begin{center}
    $q$\end{center}& \begin{center}$n$\end{center} &\begin{center}$d$\end{center} &\begin{center} Constant weight of $\mathbf{w}$\end{center} &\begin{center}Previous upper bound
\end{center}
&\begin{center}
\quad \quad Our upper bound
\end{center}
\\
\hline
5& 10& 17&2&  3 &*2 \\
6&8&14 &1& 7 &*6\\
6&9&20 &1& 3 &*2\\
17&3&18&1&3&*2\\
17&3&19&1&3&*2\\
17&4&19&2&11&8\\
17&4&20&2&8&6\\
17&4&21&2&6&4\\
17&4&23&2&3&*2\\
17&4&24&2&3&*2\\
17&5&23&2&15&12\\
17&5&24&2&11&8\\
17&5&25&2&8&6\\
17&5&26&2&6&4\\
17&5&27&2&5&4\\
17&5&29&2&3&*2\\
17&5&30&2&3&*2\\
17&5&31&2&3&*2\\
17&6&27&2&20&18\\
17&6&28&2&14&10\\
17&6&29&2&10&8\\
17&6&30&2&7&6\\
17&6&31&2&6&4\\
\hline
\end{tabular}\label{leetable}
\vspace*{0.3cm}
\\
The $*$ indicates a tight bound.
\begin{appendices}
\section{Plotkin Bound Approach}\label{secA1}
Here, we show that Theorem \ref{plot} is indeed a Plotkin-type bound. Consider we wrote all the assigned addresses of each vertex as a row of a $0,1$ matrix. Hence, we have a $n\times c_{\lambda}(K_n,\mathbf{w})$ binary matirx. Now, we count these addresses in two ways: columns and rows of the matrix. First, we can conclude that: every two addresses have at least $\lambda.w_{ij}$ distance, and we have $\binom{n}{2}$ pair addresses. Also, in the $j$-th column,  let $z_j$ be the number of $0$'s in that column. Then, the number of $1$'s will be $n-z_j$. So, each $0$ with each $1$ makes one distance and the maximum value is $\lfloor \frac{n^2}{4}\rfloor$. Therefore, we will have $\frac{c_{\lambda}(K_n,\mathbf{w})}{\lambda}\ge\frac{\sum w_{ij}}{\lfloor \frac{n^2}{4}\rfloor}$. As a result, we have for every connected weighted graph $(G,\mathbf{w})$ with $n$ vertices, $\beta(G, \mathbf{w}) \geq \frac{\sum w_{ij}}{\lfloor \frac{n^2}{4}\rfloor}$.
\section{Table of Notations}\label{secB1}
\begin{tabular}{l|l}
\textbf{Notation:} & \textbf{Definition:}  \\
$\di_\mathbf{w}(G)$ & Diameter of graph $(G,\mathbf{w})$\\
$G \square H$ & Cartesian product of graph $G$ and $H$\\
$(G,\mathbf{w})$ or $G_\mathbf{w}$ & Graph $G$ with weight function $\mathbf{w}$  \\
$h_p(G_\mathbf{w}, \mathbf{w})$ & The lowest weight of Hamilton path of $G$ \\
$h_c(G_\mathbf{w}, \mathbf{w})$ & The lowest weight of Hamilton cycle of $G$ \\
$c(G,\mathbf{w})$ &Min $m$ s.t. $\exists f:V\rightarrow\{0,1\}^m$,$\forall u,v \in V$,$d_G (u,v)\geq d_H\big(f(u), f(v)\big)$\\
$$& such $f$ is called binary addressing \\
$$& and $f(u)$ is the address of vertex $u$.\\
$c_{\lambda}(G,\mathbf{w})$ & $c(G, \lambda .\mathbf{w} )$ \\
$\beta(G, \mathbf{w}) $ & $\lim_{\lambda \rightarrow \infty}\frac{c_{\lambda}(G,\mathbf{w})}{\lambda}$ \\

\end{tabular}\label{notationtable}
\section{Table of Comparison of Problems}\label{secC1}
\scriptsize
\setlength{\arrayrulewidth}{0.3mm}
\setlength{\tabcolsep}{17pt}
\renewcommand{\arraystretch}{1.3}
\begin{tabular}{ |p{1.7cm}|p{1cm}|p{1.9cm}|p{1.5cm}|p{1.2cm}| }
\hline
\multicolumn{5}{|c|} {$f:V (G) \rightarrow s^m$}\\
 \hline\hline
Problem name& Alphabet $s$, graph G& Condition& Objective& Introduced 
\\
\hline
Addressing Problem&$\{0,1,*\}$,U& $d(f(u),f(v))=d(u,v)$&Minimizing $m$& Graham \& Pollak\cite{grah} \\
\hline
Isometric Hypercube Embedding&$\{0,1\}$,U& $d\big(f(u),f(v)\big)=d(u,v)$&Existence of embedding& Firsov \cite{iso}  \\
\hline
$\lambda$-Scale Embedding into $Q_n$&$\{0,1\}$,U& $d\big(f(u),f(v)\big)=\lambda d(u,v)$& Existence of embedding& Shpectorov \cite{scale}\\
\hline
Bi-Lipschitz Embedding&Metric spaces, U&$d(u,v)\le d\big(f(u),f(v)\big)\le cd(u,v)$&Existence of embedding& Bourgain \cite{logn}\\
\hline
BSEP&$\{0,1\}$,W&$d\big(f(u),f(v)\big)\ge d(u,v)$&Minimizing $m$& This paper\\
\hline
 $\lambda$-BSEP& $\{0,1\}$,W& $d\big(f(u),f(v)\big)\ge \lambda d(u,v)$& Value of $\beta(G,\mathbf{w})$& This paper\\
 \hline
\end{tabular}\label{comparetable}
\vspace*{0.3cm}
\\
U stands for unweighted graphs, and W for weighted ones. Also, $Q_n$ is the hypercube.
\end{appendices}
\bibliography{sn-article}
\end{document}